\documentclass[12pt]{iopart}
\usepackage{graphicx}
\usepackage{bm}
\usepackage{iopams}
\usepackage{braket}

\usepackage{changes}
\colorlet{Changes@Color}{red}

\def\app#1#2{%
  \mathrel{%
    \setbox0=\hbox{$#1\sim$}%
    \setbox2=\hbox{%
      \rlap{\hbox{$#1\propto$}}%
      \lower1.1\ht0\box0%
    }%
    \raise0.25\ht2\box2%
  }%
}

\renewcommand{\contentsname}{Dynamical modeling of pulsed two-photon interference}

\begin{document}

\title{Dynamical modeling of pulsed two-photon interference}

\author{Kevin A. Fischer}
\address{E. L. Ginzton Laboratory, Stanford University, Stanford CA 94305, USA}
\ead{kevinf@stanford.edu}
\author{Kai M\"uller}
\address{Walter Schottky Institut, Technische Universit\"at M\"unchen, 85748 Garching bei M\"unchen, Germany}
\author{Konstantinos G. Lagoudakis}
\address{E. L. Ginzton Laboratory, Stanford University, Stanford CA 94305, USA}
\author{Jelena Vu\v{c}kovi\'c}
\address{E. L. Ginzton Laboratory, Stanford University, Stanford CA 94305, USA}

\date{\today}

\begin{abstract}
Single-photon sources are at the heart of quantum-optical networks, with their uniquely quantum emission and phenomenon of two-photon interference allowing for the generation and transfer of nonclassical states. Although a few analytical methods have been briefly investigated for describing pulsed single-photon sources, these methods apply only to either perfectly ideal or at least extremely idealized sources. Here, we present the first complete picture of pulsed single-photon sources by elaborating how to numerically and fully characterize non-ideal single-photon sources operating in a pulsed regime. In order to achieve this result, we make the connection between quantum Monte--Carlo simulations, experimental characterizations, and an extended form of the quantum regression theorem. We elaborate on how an ideal pulsed single-photon source is connected to its photocount distribution and its measured degree of second- and first-order optical coherence. By doing so, we provide a description of the relationship between instantaneous source correlations and the typical experimental interferometers (Hanbury--Brown and Twiss, Hong--Ou--Mandel, and Mach--Zehnder) used to characterize such sources. Then, we use these techniques to explore several prototypical quantum systems and their non-ideal behaviors. As an example numerical result, we show that for the most popular single-photon source---a resonantly excited two-level system---its error probability is directly related to its excitation pulse length. We believe that the intuition gained from these representative systems and characters can be used to interpret future results with more complicated source Hamiltonians and behaviors. Finally, we have thoroughly documented our simulation methods with contributions to the Quantum Optics Toolbox in Python (QuTiP) in order to make our work easily accessible to other scientists and engineers.
\end{abstract}

\maketitle

\tableofcontents

\newpage


\section{Introduction}

The development of the quantum single-photon source has ushered in the field of optical quantum information technology~\cite{Eisaman2011-zm}. Such sources, serving as generators of flying photonic qubits~\cite{Reiserer2014-xm}, lie at the heart of nearly every quantum-optical technology~\cite{OBrien2009-dq}, including photonic logic gates~\cite{OBrien2003-ey}, quantum networking~\cite{Kuhn2002-lr}, and highly nonclassical NOON state generation~\cite{Matthews2011-zo}. Critical to the usefulness of a single-photon source is knowledge of its temporal profile~\cite{Woolley2013-sz, Loudon1989-nj, Legero2003-pq}, which has been accomplished via two means: heralding of the emission through a projective measurement on a continuous entangled pair source~\cite{Qin2015-fb} or on-demand generation through pulsed excitation of a system that releases only one photon per pulse~\cite{Santori2002-hd}. In this paper, we thoroughly discuss the properties of on-demand pulsed single-photon sources.

Before we begin with our detailed analysis, we provide a brief summary of the characteristics of ideal single-photon sources. Generally, a single-photon source creates a pulse containing at most one photon. More specifically, ideal and on-demand single-photon sources are characterized by their:

\begin{itemize}

\item \textit{Photocount distribution}, yielding at most one photon per pulse and hence zero \textit{second-order optical coherence}~\cite{Eisaman2011-zm, Paul1982-pg},

\item \textit{First-order optical coherence}, such that the output pulse is a pure state of the free radiation field rather than an incoherent mixture of single-photon pulses~\cite{Woolley2013-sz, Loudon1989-nj, Legero2003-pq},

\item \textit{Spatio-temporal profile}, whose precise shape is tailored by the generating system and usefulness determined by its application~\cite{Woolley2013-sz, Loudon1989-nj, Santori2002-hd}. The spatial component includes both the physical mode shape as well as its polarization.

\end{itemize}

Given that single-photon pulses exist as wavepackets, they may also display interference, and hence their spatio-temporal profiles are critical in this respect: From this realization came one of the great discoveries of the 20th century, the concept of an identical photon~\cite{Loudon1989-nj}. When a pair of photonic wavepackets share all three of the aforementioned characteristics, they are said to be identical or indistinguishable.  Meanwhile, the identical nature of photons has provided quantum optics with a rich class of experiments to explore. Perhaps the most famous is two-photon interference (seen in the Hong--Ou--Mandel interferometer), whereby two identical photons incident on a beamsplitter always exit a chosen output port together. First demonstrated in 1987 with photons produced via parametric downconversion~\cite{Hong1987-lv}, two-photon interference has been observed with photons generated via trapped ions~\cite{Maunz2007-xz}, artificial atoms~\cite{Santori2002-hd}, and even circuit QED systems~\cite{Lang2013-qv}.

While writing down the wavefunction for such photon pulses is now fairly well-established (at least from a quantum-field theoretic point of view~\cite{Blow1990-qn}), simulating non-ideal single-photon sources has proven extremely challenging. Only a few papers ever attempted to tackle this type of problem, and even so they investigated highly restrictive cases~\cite{Florjanczyk1987-bm, Santori2009-qe}. Here, we present a more general simulation technique that allows for the quantification of single-photon emission from systems characterized by Hamiltonians with nearly arbitrary time-dependence. This work also includes techniques for directly calculating the effective pulse-wise two-photon interference in both a Hong--Ou--Mandel interferometer~\cite{Hong1987-lv, Maunz2007-xz} and the often misunderstood unbalanced Mach--Zehnder interferometer~\cite{Santori2002-hd, Muller2015-vi}. Importantly, we have worked to make this technique easily accessible to experimentalists wishing to model their on-demand sources by contributing to the underlying code of the open-source package Quantum Toolbox in Python (QuTiP)~\cite{Johansson20121760}. In addition, we have provided detailed examples to the QuTiP repository demonstrating many of the simulations discussed in this paper.


\section{Prototypical single-photon sources \label{systems}}

\begin{figure*}
\begin{indented}
\item\includegraphics{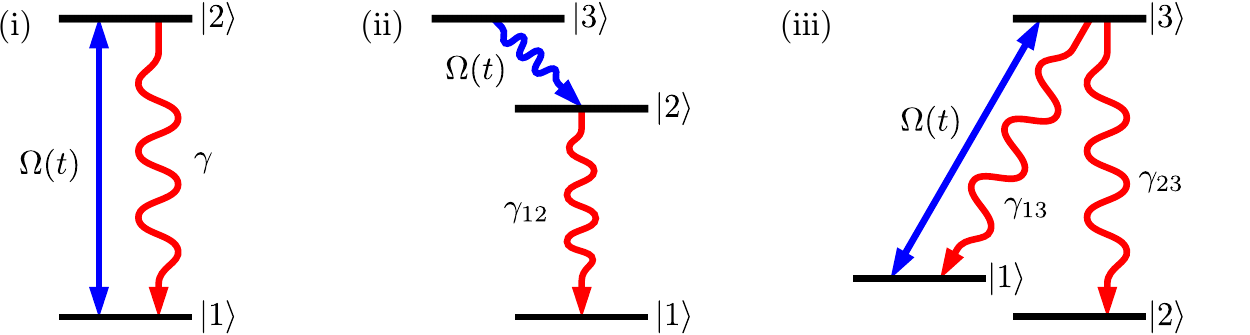}
\end{indented}
\caption{Idealized models for three single-photon sources, showing a representative variety of behaviors. (i) \textit{Coherently excited two-level system}: chosen to illustrate performance degradation primarily in its \textit{second}-order coherence. (ii)  \textit{Incoherently excited three-level ladder system}: chosen to illustrate performance degradation only in its \textit{first}-order coherence. (iii) \textit{Coherently excited three-level lambda system}:  chosen to illustrate lack of performance degradation in its ideal limit of $\gamma_{13}\rightarrow0$.}
\label{figure:1}
\end{figure*}

Although the technique we present is fairly general, we have chosen to discuss only simplistic model systems that can easily be understood and run by a broad range of quantum scientists. Similarly, while we study our systems only under drive by Gaussian pulses, our technique is trivially extensible to more complicated pulse shapes and system Hamiltonians. (For instance, see our work on single-photon emission from Jaynes--Cummings type systems~\cite{Muller2015-vi, Muller2015-il, Muller2015-om}.) Instead in this paper, our three choices of model systems (figure~\ref{figure:1}) will capture various behaviors of single-photon sources with regard to their coherences, as a function of driving pulse length. These models have been chosen as a somewhat representative set of possible on-demand single-photon sources.
\begin{enumerate}

\item \textit{Coherently excited two-level system}: This source was chosen to illustrate performance degradation primarily in its \textit{second}-order coherence with increasing pulse length relative to its spontaneous emission rate. This system is described by the following Hamiltonian~\cite{Steck2014-bk}
\begin{equation}
H_1=\hbar (\omega_0-\omega_d) \sigma^\dagger \sigma + \frac{\hbar\Omega(t)}{2}\left( \sigma\textrm{e}^{i(\omega_0-\omega_d)t} + \sigma^\dagger \textrm{e}^{-i(\omega_0-\omega_d)t}\right)\textrm{,}
\end{equation}
where $\omega_0$ is the transition frequency, $\omega_d$ the laser frequency, $\Omega(t)$ the pulse driving strength, and $\sigma$ the atomic lowering operator. We will consider the dynamics of this system with only one collapse operator, corresponding to spontaneous emission at a rate $\gamma$, i.e. $C=\sqrt{\gamma}\sigma$~\cite{Steck2014-bk}. Example experimental systems of this type can be found in coherently excited quantum dots~\cite{He2013-vb, Dada2016-jg}, trapped ions~\cite{Maunz2007-xz}, and circuit QED platforms~\cite{Lang2013-qv}.

\item \textit{Incoherently excited three-level ladder system}: This source was chosen to model performance degradation only in its \textit{first}-order coherence with increasing pulse length relative to its spontaneous emission rate. First, we assume that the system is initialized into the state $\ket{3}$ each emission cycle. From there, we will simulate only dissipative evolution through a cascade driven by two collapse operators of rates~$\Omega(t)$~and~$\gamma_{12}$, i.e.~${C_n=\{\Omega(t)\sigma_{23}, \sqrt{\gamma_{12}}\sigma_{12}\}}$.  Example experimental systems of this type can be found in electrically injected or incoherently excited quantum dots~\cite{Yuan2001-uf} and the polariton-phonon cascades of solid-state Jaynes--Cummings-like systems~\cite{Muller2015-om}.

\item \textit{Coherently excited three-level lambda system}: This source was chosen to illustrate lack of performance degradation in its ideal limit. Specifically, it always has the perfect values of its \textit{first}- and \textit{second}-order coherences for a single-photon source. Therefore, such a source can generate single-photons with nearly arbitrary wavefunctions. This system is described by the following Hamiltonian
\begin{equation}
H_3=\hbar (\omega_0-\omega_d) \sigma^\dagger_{13} \sigma_{13} + \frac{\hbar\Omega(t)}{2}\left( \sigma_{13}\textrm{e}^{i(\omega_0-\omega_d)t} + \sigma^\dagger_{13} \textrm{e}^{-i(\omega_0-\omega_d)t}\right)\textrm{,}
\end{equation}
where $\omega_0$ is the transition frequency, $\omega_d$ the laser frequency, $\Omega(t)$ the pulse driving strength, and $\sigma_{13}$ an atomic lowering operator. We will consider the dynamics of this system after having been initialized each cycle to the state $\ket{1}$ and under the influence of two collapse operators, corresponding to spontaneous emission from the excited state towards the system's two ground states. These give rise to dynamics governed by two collapse operators with rates $\gamma_{13}$ and $\gamma_{23}$, i.e.~${C_n=\{\sqrt{\gamma_{13}}\sigma_{13}, \sqrt{\gamma_{23}}\sigma_{23}\}}$. To study the ideal case of arbitrary single-photon generation, we will set $\gamma_{13}\rightarrow0$.  Example experimental systems of this type can be found in trapped ions~\cite{Kuhn2002-lr, Keller2004-vb}.
\end{enumerate}
For each of these systems, its first-order coherence can be decreased by the effect of any pure dephasing terms as well. These terms are modeled by the inclusion of collapse operators with the form $C_n=\sqrt{\gamma_{d,n}(t)}\sigma_{nn}$, which may potentially be time-dependent.

Finally, we note that it is trivial to model coupling to different spatial modes by decomposing a given collapse operator into independent collapse operators. For example, consider the replacement where $C=\sqrt{\gamma}\sigma\rightarrow C_n=\{\sqrt{\gamma/2}\sigma,\sqrt{\gamma/2}\sigma\}$: Then the emission occurs at the same total rate but into two separate spatial modes (with potentially different polarizations). Because this phenomenon is well understood~\cite{Woolley2013-sz}, we will implicitly assume that all of our sources emit into the same spatial mode for this paper.


\section{Modeling of source dynamics\label{sec:sourceDynamics}}

The dynamics of the above systems are governed by the quantum-optical master equation~\cite{Steck2014-bk, Carmichael2009-uc}
\begin{equation}
\frac{\mathop{\textrm{d}}}{\mathop{\textrm{d} t}}\rho(t)=\mathcal{L}(t)\rho(t)\textrm{,}
\end{equation}
where $\mathcal{L}(t)$ is the Liouvillian super-operator that characterizes the time-dependent system evolution. More specifically, in the Markovian limit of system-reservoir interactions
\begin{equation}
\hspace{-20pt}
\frac{\mathop{\textrm{d}}}{\mathop{\textrm{d} t}}\rho(t)=-\frac{i}{\hbar} \left[ H(t), \rho(t) \right]+\sum_n\frac{1}{2}\left[ 2C_n\rho(t)C_n^\dagger-\rho(t)C_n^\dagger C_n-C_n^\dagger C_n\rho(t) \right]\textrm{,}
\end{equation}
where $C_n=\sqrt{\gamma_n}a_n$ are the collapse operators referenced above ($a_n$ are the operators through which the system couples to the environmental modes). As discussed in the introduction, single-photon sources are characterized by potentially non-trivial first- and second-order optical coherences. Therefore, we will be interested to use the system dynamics to calculate correlations of the form
\begin{equation}
G(t,\tau)=\langle A(t) B(t+\tau)C(t)\rangle\textrm{,}
\end{equation}
where $A$, $B$, and $C$ are each some combination of $a_n$, $a^\dagger_n$, and the identity matrix. Although this calculation is often performed with the quantum regression theorem~\cite{Gardiner1991-pe}, its formal statement excludes time-dependent Liouvillians in all references to the authors' knowledge. Yet, when studying a system under pulsed excitation, a time-dependent Liouvillian invariably arises due to the dynamical driving. Therefore, we have extended the quantum regression theorem to time-dependent Liouvillians below.

When the quantum regression theorem is applied for time-dependent Liouvillians, it inherits the approximations from the quantum-optical master equation and yields the following result
\begin{equation}
\langle A(t) B(t+\tau)C(t)\rangle = \textrm{Tr}_{\textrm{sys}}\{B \Lambda (t, t+\tau) \}\textrm{,}
\end{equation}
where $\Lambda(t, t+\tau)$ is governed by the evolution equation 
\begin{equation}
\frac{\partial}{\partial\tau} \Lambda(t,t+\tau) = \mathcal{L}(t+\tau) \Lambda(t,t+\tau) 
\end{equation}
and is subject to the initial condition
\begin{equation}
\Lambda(t,t) = C\rho(t)A\textrm{.}
\end{equation}
Although this equation only can evolve forward in time, the correlators discussed in this paper either give rise to physical measurements and inherently require $\tau>0$~\cite{Cresser1987-qg} or possess a conjugate time-reversal symmetry about $\tau$~\cite{Steck2014-bk}. We note that the authors have added this algorithm as a routine to QuTiP, so a simple call to the correlators automatically calculates two-time correlations with time-dependent Liouvillians.

This evolution equation has a particularly nice interpretation when the probability for three successive photodetection events is negligible and
\begin{equation}
\langle A(t) B(t+\tau)C(t)\rangle=\langle a_n^\dagger(t) a_n^\dagger(t+\tau)a_n(t+\tau)a_n(t)\rangle\textrm{,}
\end{equation}
which then represents the probability density of detecting an excitation at time~$t$ followed by another excitation at time $t+\tau$. The quantum regression theorem clearly captures this through: first a reduction of the density of matrix by one excitation
\begin{equation}
\Lambda(t,t) = a_n\rho(t)a_n^\dagger
\end{equation}
and by then computing the probability density of a second reduction
\begin{equation}
\textrm{Tr}_{\textrm{sys}}\{a_n^\dagger a_n \Lambda(t,t+\tau)\}\textrm{.}
\end{equation}

Finally, we comment on the connection between the system correlators and the free radiation-field correlators. Because single-photon detectors make measurements on the free radiation field, we physically measure correlations having to do with the field-flux operators, e.g. $b(t)$. Fortunately, Gardiner and Zoller's input--output theory~\cite{Gardiner1991-pe} provides a direct connection between the internal system operators and external radiation mode operators, such that measuring flux correlations is equivalent to measuring the internal system correlators. More specifically, if a system operator $a_n(t)$ is coupled to the external radiation modes represented by $b(t)$ and we wish to calculate a physical correlator involving $a_n(t)$, then we may simply make the replacement $b(t)\rightarrow\sqrt{\gamma_n} a_n(t)$.


\section{Single-photon source photocount distribution}

As discussed in the introduction, an ideal pulsed single-photon source would contain only a single-photon per pulse and hence only a single quanta of energy in its wavepacket---in this section we will fully explore this concept. Consider photon flux incident on a photon counter with finite timing resolution. Mathematically, the probability that the flux results in a detection event between $t$ and $t+\mathop{\textrm{d} t}$ is
\begin{equation}
p(t) \mathop{\textrm{d} t} = \eta \langle \hat{f}(t) \rangle \mathop{\textrm{d} t}\textrm{,}
\end{equation}
where $\eta$ is the total detection and collection efficiency and $\hat{f}(t)$ is the instantaneous photon flux operator (analogous to classical intensity)~\cite{Loudon2000-li}. Notably, the instantaneous detection probability $\eta \langle \hat{f}(t) \rangle$ must be negligible such that multiple photoionizations in the detector cannot occur.  Integrating over the photon pulse duration, $T$, there exists a classical probability distribution $P_m(T)$ that governs the number of expected photocounts. A single-photon source requires
\begin{equation}
\label{cases}
P_m (T) = \cases{
1-\eta&if $m=0$\\
\eta&if $m=1$\\
0&otherwise\\}\textrm{,}
\end{equation}
where $m\in\{\mathbb{Z}\ge0\}$ is a classical random variable that represents the number of photocounts. 

Thus, it is sufficient to characterize a single-photon source by a measure which roughly corresponds to the probability of two events occurring within $T$: the second-order factorial moment of its photocount distribution~\cite{Carmichael2009-uc}
\begin{equation}
G^{(2)}[0]\equiv\textrm{E}\left[ m(m-1) \right]\textrm{.}
\end{equation}
(Because the photocount distribution is over the classical random variable $m$, we use the $\textrm{E}\left[ ... \right]$ notation to denote the classical expectation value.) While the first detection still depends on $m$, the second detection depends on $m-1$ because the first detection subtracts a quantum of energy from the distribution~\cite{Loudon2000-li}. Therefore, for an ideal pulsed single-photon source  $G^{(2)}[0]=0$.

Unfortunately, the detection and collection efficiency $\eta$ can be quite difficult to experimentally measure, which makes this correlation challenging to directly estimate. Instead we will focus on determining its normalized version
\begin{equation}
g^{(2)}[0] \equiv \frac{G^{(2)}[0]}{\textrm{E}\left[ m \right]^2} = \frac{\textrm{E}\left[ m(m-1) \right]}{\textrm{E}\left[ m \right]^2}\textrm{.}
\end{equation}
This normalized second-order factorial moment of $P_m (T)$ is referred to as the \textit{measured degree of second-order coherence at zero time delay} and doesn't depend on the efficiency~$\eta$~\cite{Loudon2000-li}. This quantity is useful because any value $g^{(2)}[0]<1$ is disallowed by classical physics~\cite{Gardiner1991-pe} and because $g^{(2)}[0]$ approximates the error probability of the source to produce two detection events relative to the number of single detection events.

Because the photocount distribution can only be estimated from the outcome of many pulse-wise experiments, a typical experimental cycle will involve periodic generation of the photon wavepacket and its subsequent detection every $t_r$ seconds. The number of photodetections from a given pulse at the time bin $nt_r$ can then be represented by the classical random variable $m[nt_r]=\{\mathbb{Z}\ge0\}$. We can then extend the pulse-wise definition of $G^{(2)}[0]$ to non-zero time delays, i.e.
\begin{equation}
G^{(2)}[k t_r] = E\left[m[0]m[kt_r]\right]
\textrm{~~for~~} k>0 \textrm{,}
\end{equation}
which we refer to as the un-normalized second-order intensity correlation at time delay~$nt_r$. We note that its time origin is irrelevant due to the quasi-stationary nature of the pulses. If the probability for three photodetections over a given pulse is negligible, then this correlation represents the probability that two photons are detected between pulses separated by the time difference~$k t_r$. From this definition, we can arrive at another definition of $g^{(2)}[0]$ which will turn out to be experimentally most useful:
\begin{equation}
g^{(2)}[0] =\frac{G^{(2)}[0]}{G^{(2)}[kt_r]}
\textrm{~~for~~} k>0 \textrm{,}
\end{equation}
provided that we have correctly chosen $t_r$ longer than the correlation time of any system operators (where $G^{(2)}[k t_r]=E[m]^2$). Experimental realities may sometimes inhibit the realization of this criterion and one can then simply take
\begin{equation}
g^{(2)}[0] =\lim_{k\rightarrow\infty}\frac{G^{(2)}[0]}{G^{(2)}[kt_r]} \textrm{.}
\end{equation}
While this definition may seem a trivial extension, this form is most useful (due to the limitations of legacy measurement instrumentation) in discussing experimental setups that estimate $g^{(2)}[0]$. Thus, we have discussed how experimental single-photon sources can be readily characterized even in the face of unknown detection and collection efficiencies.


\subsection{Hanbury--Brown and Twiss interferometer}

\begin{figure*}
\begin{indented}
\item\includegraphics{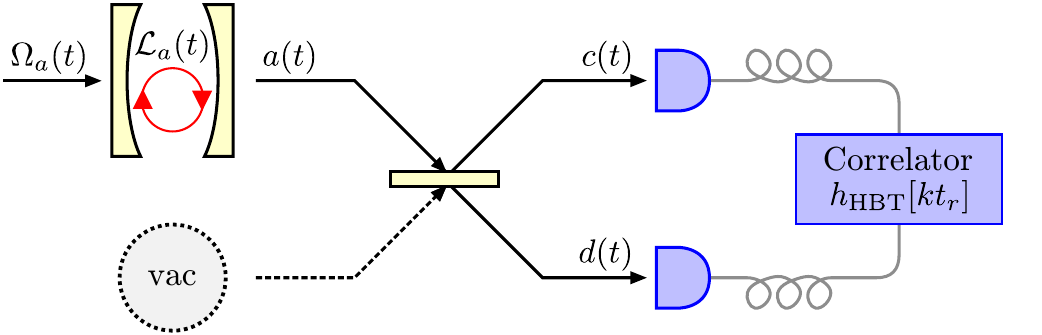}
\end{indented}
\caption{Schematic of the \textit{Hanbury--Brown and Twiss interferometer}: The source's emission (periodic every $t_r$) is path-entangled by a beamsplitter and measured by two detectors. A digital recorder then correlates the detection times and computes $h_{\scriptsize\textrm{HBT}}[kt_r]$. Here, $\Omega_a(t)$ indicates the coherent driving field, $\mathcal{L}_a(t)$ represents the dynamics of the source, $a(t)$ represents the output of the source, and $c(t)$ and $d(t)$ represent the fields at the inputs to the detectors.}
\label{figure:2}
\end{figure*}

While one can easily estimate $g^{(2)}[0]$ from the detection record of ideal photon counters, this is not possible for most experimental detectors and single-photon sources. Experimental photodetectors have a so-called ``dead time'', for which the detector cannot register a second count following the first, that usually is much longer than the temporal length of the photon pulse. With a maximum of one registered count per pulse, such a detector could never distinguish between single- and multi-photon sources. Fortunately, it turns out that under the right approximations, an experimental setup known as the Hanbury--Brown and Twiss interferometer (figure~\ref{figure:2}) is still capable of precisely estimating $g^{(2)}[0]$ by using two photon counters, and hence is still capable of characterizing realistic single-photon sources~\cite{Koashi1993-zh}.

Because our focus is on pulsed single-photon sources, we will only discuss the Hanbury--Brown and Twiss (HBT) interferometer operated in a pulsed mode of operation, where we repeatedly excite our source every $t_r$ seconds~\cite{Koashi1993-zh}. In the HBT setup, every pulse is path-entangled between two channels by a $50:50$ beamsplitter and each channel is fed into a single-photon detector. The periodic photon absorption events registered by the two detectors can be represented as the classical random variables $m_c[nt_r]=\{0,1\}$ and $m_d[nt_r]=\{0,1\}$ where $n\in\{\mathbb{Z}>0\}$; They may take on unity values at times $nt_r$ to represent photon detections. Here, the square brackets will indicate the discrete nature of the random variables and their intensity correlations, due to both their periodic pulsed nature and any timing uncertainty in their detection (either explicit by detector jitter or implicit through purposeful erasure of timing information). Although an estimate for $P_m(T)$ could be built up through many periodic trials and used to directly compute $g^{(2)}[0]$, traditionally  $g^{(2)}[0]$ has been extracted from a histogram of the time-correlated detection records, i.e.
\begin{equation}
h_{\scriptsize\textrm{HBT}}[kt_r]=\sum_{n=0}^{\lfloor\frac{T_{\scriptsize\textrm{int}}}{t_r}\rfloor} m_c[nt_r] m_d[(n+k)t_r]\textrm{,}
\end{equation}
where $k\in\mathbb{Z}$, $T_{\scriptsize\textrm{int}}$ is the integration time, and $kt_r$ is the time difference between detections~\cite{Santori2003-dk}. Each histogrammed time-bin $h_{\scriptsize\textrm{HBT}}[kt_r]$ is an independent and binomially-distributed random variable, whose standard deviation estimator is given by $\sqrt{h_{\scriptsize\textrm{HBT}}[kt_r\left](1-\eta\right)}\approx\sqrt{h_{\scriptsize\textrm{HBT}}[kt_r]}$ for most experiments ($\eta$ is again the combined detection and collection efficiency).

After careful consideration of the detector nonidealities such as dead time and dark counts, as well as of the expected statistics of a single-photon source~\cite{Santori2003-dk}, one can arrive at the approximation
\begin{equation}
\textrm{E}\left[h_{\scriptsize\textrm{HBT}}[kt_r]\right]\approx \frac{\eta^2}{4} \Big\lfloor\frac{T_{\scriptsize\textrm{int}}}{t_r}\Big\rfloor G^{(2)}[kt_r]\textrm{.}
\end{equation}
Here, the factor of $\frac{1}{4}$ accounts for the action of the beamsplitter to halve the signal at each detector. In making the connection back to the second-order factorial moment at zero delay that was discussed in the previous section, consider
\begin{equation}
G^{(2)}[0]\approx\frac{4}{\eta^2}\frac{1}{\lfloor\frac{T_{\scriptsize\textrm{int}}}{t_r}\rfloor}\sum_{n=0}^{\lfloor\frac{T_{\scriptsize\textrm{int}}}{t_r}\rfloor}\textrm{E}\left[ m_c[nt_r] m_d[nt_r]\right]\textrm{.}
\end{equation}
While it may be surprising that this expression is equivalent to $G^{(2)}[0]$, i.e. $\textrm{E}\left[ m(m-1) \right]$, the key insight is that the random variables $m_c[nt_r]$ and $m_d[nt_r]$ are not independent because a detection event by either detector pulls one quantum of energy from the total path-entangled field.

\begin{figure}
\begin{indented}
\item\includegraphics{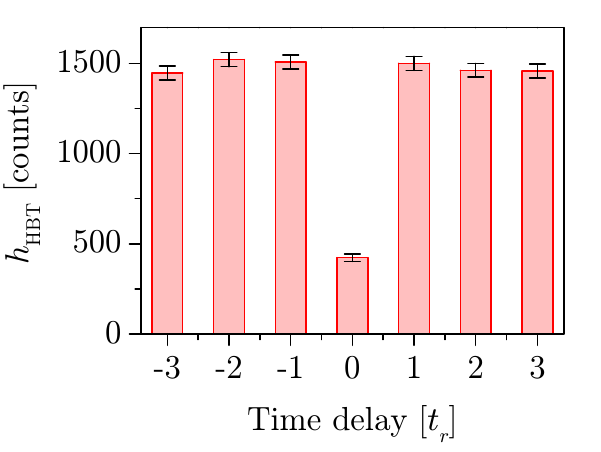}
\end{indented}
\caption{Example Hanbury--Brown and Twiss histogram generated using pulsed single-photon emission from a quantum-dot-based cavity quantum electrodynamical system. An estimate for the measured degree of second-order coherence can be obtained by taking the ratio of the counts at zero delay, $h_{\scriptsize\textrm{HBT}}[0]$, to the counts at delay $t_r$, $h_{\scriptsize\textrm{HBT}}[t_r]$. This appropriately estimates $g^{(2)}[0]$ because the pulse-wise correlations have already disappeared after one repetition time.}
\label{figure:3}
\end{figure}

As previously mentioned, however, it is difficult to experimentally estimate $G^{(2)}[0]$ due to unknown setup efficiencies. Fortunately, since all correlations in $G^{(2)}[kt_r]$ are lost at long times such that 
\begin{equation}
\lim_{k\rightarrow\infty} G^{(2)}[kt_r]=\textrm{E}\left[  m\right]^2\textrm{,}
\end{equation}
then $\lim_{k\rightarrow\infty}\textrm{E}\left[h_{\scriptsize\textrm{HBT}}[kt_r]\right]$ can serve as an intensity reference. Thus, we can obtain an estimate for the measured degree of second-order coherence from the following ratio
\begin{equation}
\hat{g}^{(2)}[0]=\lim_{k\rightarrow\infty}\frac{h_{\scriptsize\textrm{HBT}}[0]}{
h_{\scriptsize\textrm{HBT}}[kt_r]}\left(1\pm\sqrt{\frac{1}{h_{\scriptsize\textrm{HBT}}[0]}+\frac{1}{h_{\scriptsize\textrm{HBT}}[kt_r]}}\right)\textrm{.}
\end{equation}
Importantly, several approximations were required to arrive at this result~\cite{Santori2003-dk}:
\begin{itemize}
\item The net detection probability per pulse $\frac{1}{2}\eta\textrm{E}\left[  m\right]$ must be very small relative to the dead time.
\item The net detection probability per pulse $\frac{1}{2}\eta\textrm{E}\left[  m\right]$ must be very small relative to the repetition time $t_r$.
\item The probability of many-photon detection must be moderately low; Analogously, the higher-order factorial moments of the photocount distribution must be of order unity, i.e. $g^{(n)}[0]\sim1$ for $n>2$.
\item The detectors' dark count rates $d_n$ are minimal, such that
\begin{equation}
G^{(2)}[kt_r]\gg \frac{1}{\eta^2}d^2+\frac{1}{\eta}\textrm{E}\left[ m \right] d\textrm{.}
\end{equation}
\end{itemize}
As an example of such a histogram, we reproduce data from reference~\cite{Muller2015-il} in figure~\ref{figure:3}. In this experiment, a single-photon source was run in pulsed mode and time-correlated using an HBT setup. Because all pulse-wise correlations have decayed already after one repetition time in this experiment,
\begin{equation}
\textrm{E}\left[h_{\scriptsize\textrm{HBT}}[t_r]\right]=\lim_{k\rightarrow\infty}\textrm{E}\left[h_{\scriptsize\textrm{HBT}}[kt_r]\right]
\end{equation}
and therefore
\begin{equation}
\hat{g}^{(2)}[0]=\frac{h_{\scriptsize\textrm{HBT}}[0]}{
h_{\scriptsize\textrm{HBT}}[t_r]}\left(1\pm\sqrt{\frac{1}{h_{\scriptsize\textrm{HBT}}[0]}+\frac{1}{h_{\scriptsize\textrm{HBT}}[t_r]}}\right)\textrm{,}
\end{equation}
which yields $\hat{g}^{(2)}[0]=0.29\pm0.04$.

Finally, we note that many experimental HBT interferometers do not exactly histogram $h_{\scriptsize\textrm{HBT}}[kt_r]$. Rather, they only approximate $h_{\scriptsize\textrm{HBT}}[kt_r]$ by electronically time-correlating detections on-the-fly. This is done by taking the first detector as the signal to start timing and the second detector as the signal to stop timing: Each start-stop sequence generates a count in the time-bin of the timer value. This way, the histogram is built up in real-time as new measurement correlations are recorded. The downside to this method is that each successively longer time-bin requires more failed detections to register a count, where the actual histogram constructed is
\begin{equation}
\hat{h}_{\scriptsize\textrm{HBT}}[kt_r]=\sum_{n=0}^{\lfloor\frac{T_{\scriptsize\textrm{int}}}{t_r}\rfloor} m_c[nt_r]  \prod_{l=0}^{k-1} (1-m_d[(n+l)t_r])m_d[(n+k)t_r]\textrm{.}
\end{equation}
Here the product term means that at long time-correlations, $\textrm{E}\left[h[kt_r]\right]\propto(1-\frac{1}{2}\eta\textrm{E}\left[  m\right])^k$ approximately and hence it decays to zero. Therefore, this electronic method of time-correlating to estimate $g^{(2)}[0]$ may not always work well if the $G^{(2)}[n t_r]$ shows correlated behavior for large $n$. Now that we have a good understanding of experimentally how to characterize the photocount distribution of a single-photon source, we will discuss the theoretical connection to the instantaneous correlations of the photon wavepacket's fields.


\subsection{Connection to correlations of instantaneous fields\label{sec:inst_corr}}

From a numerical modeling perspective, estimating $g^{(2)}[0]$ for a given system Liouvillian is fairly straightforward using Monte--Carlo wavefunction techniques~\cite{Molmer1993-af}. Here, evolution of the system wavefunction is modeled, conditioned on the detection events of an ideal single-photon detector with infinite bandwidth. Such a detector may absorb one quantum of energy from the emitted field at a time, in proportion to the photon flux at a given instant, and can distinguish detection events with infinite timing resolution.  Through this process, detection records are generated that build up an estimate for $P_m(T)$ and hence $g^{(2)}[0]$. However, the Monte--Carlo method often requires an extremely large number of simulated evolutions (trajectories) to obtain an acceptable approximation of the measured degree of second-order coherence.

Fortunately, for systems with reasonably-dimensioned Hamiltonians, there exists a faster and more intellectually satisfying way of computing $g^{(2)}[0]$---as used for pulsed nonclassical light sources, this algorithm was first implemented numerically in our previous work~\cite{Muller2015-il}. This method directly relies on using the quantum regression theorem, as outlined in section \ref{sec:sourceDynamics}, to compute correlations between the instantaneous system operators~$a_n(t)$ previously discussed. These correlations in turn are related to the instantaneous correlations of the continuous-mode free-field operator $b(t)$ that describes the emitted photon-field flux. In this section, we now fully elaborate on this explicit connection of the measured degree of coherence with its associated instantaneous field correlations.

We begin with a more directly quantum mechanical model of our ideal single-photon detector that formalizes and combines the stories from references~\cite{Blow1990-qn, Loudon2000-li} and~\cite{Koashi1993-zh}. In analogy to the classical integrated mean intensity, the quantum mechanical operator
\begin{equation}
\hat{M} (T) =\int_0^T \mathop{\textrm{d} t} \, \hat{f}(t)=\int_0^T \mathop{\textrm{d} t} \, b^{\dagger}(t) b(t)
\end{equation}
represents the total photon number arriving at an ideal detector over the time interval $t\in[0,T]$ (where $T$ is again over the duration of the photon pulse). As such, its quantum mechanical expectation value yields $\langle \hat{M}(T) \rangle=\textrm{E}\left[  m\right]$. Comparing our semiclassical definition of $G^{(2)}[0]$ to the quantum mechanical operator $\hat{M}(T)$ we have
\begin{equation}
G^{(2)}[0] = \langle \hat{M}(T)(\hat{M}(T)-1) \rangle 
\textrm{~~~~~or~~~~~}
G^{(2)}[0] =\langle: \hat{M}(T)^2:\rangle\textrm{.}
\end{equation}
Here, $\langle: \hat{M}(T)^2 :\rangle$ denotes the quantum mechanical expectation value of the normally-ordered second moment of the photon number operator $\hat{M}(T)$. Writing out this moment explicitly
\begin{equation}
G^{(2)}[0] =\int_0^T \int_0^T \mathop{\textrm{d} t} \mathop{\textrm{d} t'} \, \langle \mathcal{T}_-[b^{\dagger}(t) b^{\dagger}(t')] \mathcal{T}_+[b(t') b(t)] \rangle\textrm{,}
\end{equation}
where the operators $\mathcal{T}_\pm$ indicate the time-ordering required of a physical measurement~\cite{Knoll1986-af} (operators with higher time indices towards the center of the expression).

We can also compute the quantum mechanical version of the normalized second-order moment $g^{(2)}[0]$, with
\begin{equation}
g^{(2)}[0] = \frac{G^{(2)}[0]}{\langle \hat{M}(T)\rangle^2}=\frac{\langle: \hat{M}(T)^2 :\rangle}{\langle \hat{M}(T)\rangle^2}
\end{equation}
and its explicit form
\begin{equation}
g^{(2)}[0] = \frac{\int_0^T \int_0^T \mathop{\textrm{d} t} \mathop{\textrm{d} t'} \, \langle \mathcal{T}_-[b^{\dagger}(t) b^{\dagger}(t')] \mathcal{T}_+[b(t') b(t)] \rangle}{(\int_0^T \mathop{\textrm{d} t} \, \langle b^{\dagger}(t) b(t) \rangle)^2}\textrm{.}
\end{equation}
Written another way by defining
\begin{equation}
G^{(2)}(t,t')\equiv\langle \mathcal{T}_-[b^{\dagger}(t) b^{\dagger}(t')] \mathcal{T}_+[b(t') b(t)] \rangle\textrm{,}
\end{equation}
then
\begin{equation}
g^{(2)}[0]= \frac{\int_0^T \int_0^T \mathop{\textrm{d} t} \mathop{\textrm{d} t'} \, G^{(2)} (t,t')}{\langle \hat{M}(T) \rangle^2}  \textrm{.}
\end{equation}

Therefore, the measured $g^{(2)}[0]$ actually represents the sum of all field-flux correlations $G^{(2)} (t,t')$ over the detection time. This expression agrees with intuition when higher-order correlations are negligible, as it then represents the probability to detect two photons at every possible pair of times, normalized to the total photon number squared. As previously discussed, input--output theory provides a direct connection between the internal system operators and external mode operators such that measuring zero delay flux correlations is equivalent to calculating the mode correlators. Thus, we may simply calculate
\begin{equation}
g^{(2)}[0] = \frac{\int_0^T \int_0^T \mathop{\textrm{d} t} \mathop{\textrm{d} t'} \, \langle \mathcal{T}_-[a_n^{\dagger}(t) a_n^{\dagger}(t')] \mathcal{T}_+[a_n(t') a_n(t)] \rangle}{(\int_0^T \mathop{\textrm{d} t} \, \langle a_n^{\dagger}(t) a_n(t) \rangle)^2}\textrm{,}
\end{equation}
where $a_n(t)$ could be an atomic lowering operator in the case of radiation from a few-level system or a cavity mode operator in the case of radiation from a cavity. Hence, we have finished outlining our novel method of numerical simulation that will be used for the rest of this paper in modeling the dynamics of pulsed single-photon sources.

Finally, we note that the above expressions will later also be expanded to directly consider correlations between two field operators labeled, for instance, $c(t)$ and $d(t)$ with
\begin{equation}
G^{(2)}_{cd}(t,t') \equiv\langle \mathcal{T}_-[c^{\dagger}(t) d^{\dagger}(t')] \mathcal{T}_+[d(t') c(t)] \rangle\textrm{,}
\end{equation}
\begin{equation}
G_{cd}^{(2)}[0] \equiv\langle: \hat{M}_c(T)\hat{M}_d(T):\rangle\textrm{,}
\end{equation}
and
\begin{equation}
g_{cd}^{(2)}[0] \equiv\frac{\langle: \hat{M}_c(T)\hat{M}_d(T):\rangle}{\langle \hat{M}_c(T)\rangle\langle \hat{M}_d(T)\rangle} \textrm{.}
\end{equation}
%


\subsection{Simulated second-order optical coherences of single-photon sources\label{sec:sim1}}

\begin{figure}
\begin{indented}
\item\includegraphics{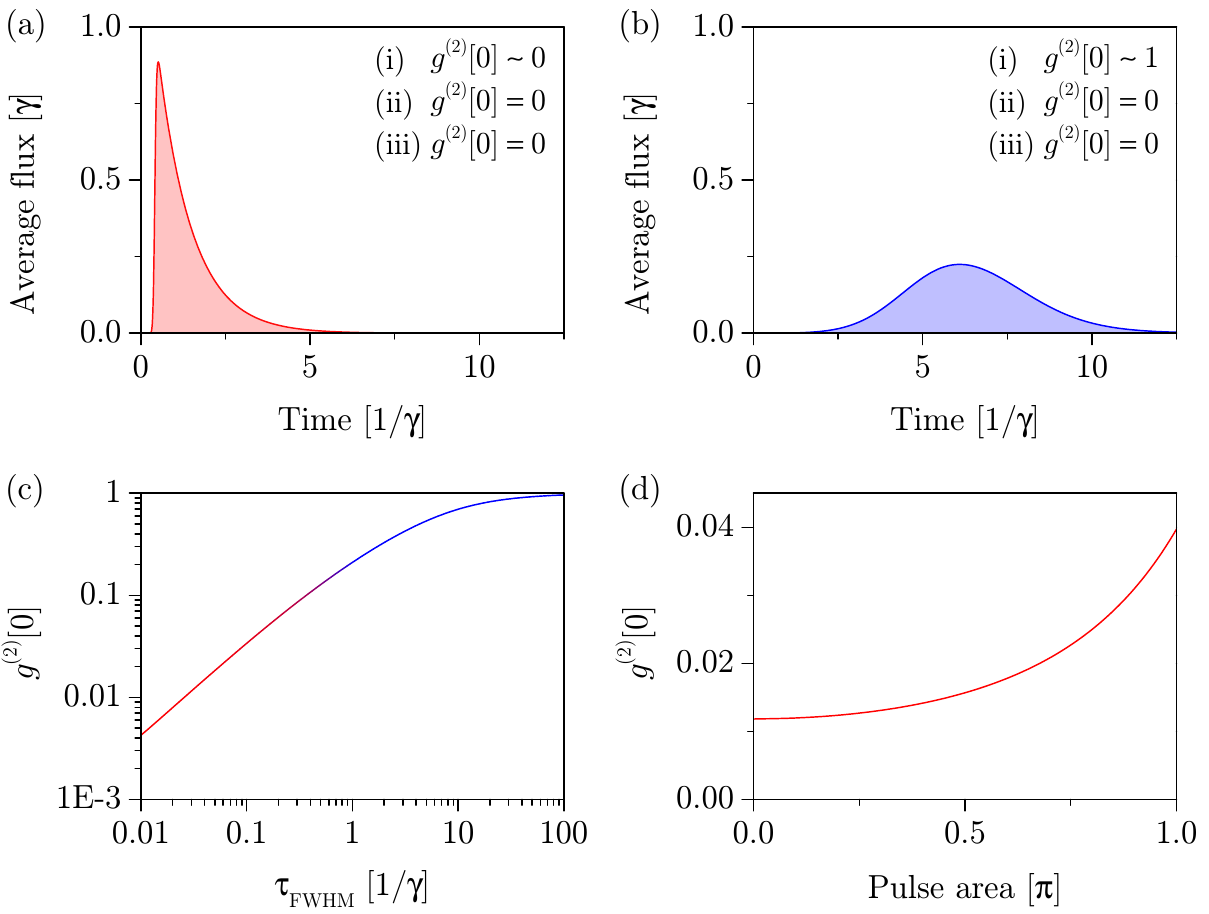}
\end{indented}
\caption{Simulations of the measured degrees of second-order coherence for the three representative systems highlighted in section~\ref{systems}, which could be experimentally measured with a Hanbury--Brown and Twiss setup. (a,b) These systems are excited by Gaussian pulses of varying pulse length, which generally results in two types of pulse envelopes, exponential (red) or Gaussian (blue). As examples, we show wavepackets of an exponential shape (a) generated with pulse length of $\tau_{\scriptsize\textrm{FWHM}}=0.1/\gamma$ in energy or a Gaussian shape (b) generated with a pulse length of $\tau_{\scriptsize\textrm{FWHM}}=3.3/\gamma$ in energy. All of the prototypical systems act as single-photon sources with zero or nearly zero $g^{(2)}[0]$ for short excitation pulses (a), but only systems (ii) and (iii) maintain this quality for long excitation pulses (b). (c) $g^{(2)}[0]$ as a function of excitation pulse length for system (i). For the simulations in (a-c), we chose the pulse length such that the average number of photons emitted is one. On the other hand in (d), $g^{(2)}[0]$ as a function of excitation pulse \textit{area} for system (i) is shown.}
\label{figure:4}
\end{figure}

Now that we have fully elaborated on the theory behind the HBT setup, we can put all of the aforementioned pieces together and begin to simulate the three different single-photon sources discussed in section~\ref{systems}. These systems were chosen because even single-photon sources with more complicated Hamiltonians and characters may potentially be mapped onto the source behaviors we will present in the following simulation sections.

The first feature we will consider is the energetic shape of the wavepackets emitted from the three systems. In fact for a given pulse, the three systems generate nearly identically shaped wavepackets when excited to emit, on average, one photon. Here, the shape of the wavepacket refers to the profile of the average energy density at a given point in space and time. Importantly, for long pulses, we opted against the standard definition~\cite{Rzazewski1984-qr} of using constant pulse area to define our pulse lengths. Instead, we chose to determine our pulse lengths such that each system emits, on average, one photon per pulse. For short pulses the two definitions agree, but for long pulses this difference allows us to focus exclusively on photon statistics for comparable average photon flux.

Using these definitions, we describe the disparate situations when the three sources are excited by temporally short or long pulses. When excited by a short Gaussian pulse, where short is relative to the systems' characteristic emission times, then the resulting wavepackets have exponentially decaying energy densities with time. An example of this behavior is shown in figure~\ref{figure:4}a (red), with a pulse length of $\tau_{\scriptsize\textrm{FWHM}}=0.1/\gamma$ in energy. On the other hand, consider the systems when excited by a long Gaussian pulse. Then the resulting wavepackets have almost Gaussian shapes, as shown in figure~\ref{figure:4}b (blue), with a pulse length of $\tau_{\scriptsize\textrm{FWHM}}=3.3/\gamma$ in energy. Wavepackets of these two shapes will be considered for the rest of the paper and their shapes will always be denoted by the colors of red (exponential) or blue (Gaussian).

Given that the energy density profiles match so closely between our prototypical sources excited with the same pulses, the differences between their wavepackets must lie elsewhere. As we will see, such differences will be seen in their quantum statistics (optical coherences). In this section, we make comparisons between their second-order optical coherences, while in the next simulation section~\ref{sec:sim2} we will compare values based on first-order optical coherences. The simulated $g^{(2)}[0]$ values for wavepackets emitted by both three-level sources, systems (ii) and (iii), are quite trivial: They are manifestly zero for all possible excitation pulses. This result is quite simple to understand---the act of initializing the system to a third state [state $|3\rangle$ for system (ii) and state $|1\rangle$ for system~(iii)] means only a single quanta of energy can leave the system before it is reinitialized. The initial and final states for both sources are unconnected so that the systems cannot be re-excited to emit multiple photons~\cite{Keller2004-vb}. These results are summarized in the legends of figures~\ref{figure:4}a~and~\ref{figure:4}b.

On the other hand, because the initial and final states are the same for the two-level single-photon source [system (i)], re-excitation may occur. Therefore, its $g^{(2)}[0]$ values exhibit a more complicated character and may be nonzero. For example, consider the effects of pulse length on $g^{(2)}[0]$ for the two-level system (shown in figure~\ref{figure:4}c). If a short excitation pulse drives photon emission, then the excitation occurs over a very short timescale as compared to the actual emission and re-excitation is very unlikely to occur. Therefore, the system possesses low second-order coherence and acts as a single-photon source (over the red side of the curve). In fact, this re-excitation probability is linear with pulse length at low powers and therefore we can fit
\begin{equation}
\hat{g}^{(2)}[0]=0.4\gamma\,\tau_{\scriptsize\textrm{FWHM}}\pm0.003\textrm{.}
\end{equation}
(Surprisingly, some papers discussing experimental results on state-of-the-art quantum dot sources quote values for $g^{(2)}[0]$ that are lower than this limit, e.g. those in reference~\cite{Schneider2016-dl}.) Because the re-excitation probability is always finite, $g^{(2)}[0]$ also has a limiting value even for arbitrarily low powers. This effect can be seen in figure~\ref{figure:4}d, where $g^{(2)}[0]$ as a function of pulse area is shown for a short pulse (pulse length of $\tau_{\scriptsize\textrm{FWHM}}=0.1/\gamma$ in energy). Therefore, even in the absence of all other non-idealities, pulsed two-level systems will never exhibit perfect single-photon emission. We note that our modeling agrees very well with recent experimental results regarding the emission from a neutral quantum dot~\cite{Dada2016-jg}.

Meanwhile, if a long excitation pulse drives photon emission from system (i), then re-excitation is highly likely and the system inherits the photon statistics of the laser pulse ($g^{(2)}[0]$=1). This regime can be seen in the blue side of figure~\ref{figure:4}c and is tabulated in figure~\ref{figure:4}b. From these simulations, we can conclude that there are regimes where a pulsed two-level system does not operate as single-photon emitter at all.


\section{Interferometers for observing pulse-wise two-photon interference}

In the previous section, we discussed how the HBT interferometer characterizes a single photon by measuring the second-order intensity interference of its path-entangled wavepacket. There the spatio-temporal profiles were irrelevant to the pulse-wise results---now we will consider interferometers whose outputs depend on these profiles. However, we will implicitly assume that the wavepackets occupy comparable spatial modes so we can focus on their temporal characteristics. Next, we will introduce several important findings and concepts that we will elaborate on and prove in the subsequent sections.

Specifically, we will discuss two interferometers that compare \textit{two} independent photon wavepackets and depend both on second-order (intensity) and first-order (field) interference: the Hong--Ou--Mandel interferometer~\cite{Woolley2013-sz} and the unbalanced and doubly-excited Mach--Zehnder interferometer~\cite{Santori2002-hd}. For these interferometers in the limit of identical single-photon inputs, the intensity correlations equal zero just like in the HBT setup. Interestingly, this means that the remaining information is encoded in the mutual first-order optical interference between wavepackets. Therefore, this correlation could in principle be measured through a traditional Michelson interferometer and is more recognizable as interference in classical optics. However, a Michelson interferometer can only measure the ensemble average of first-order coherence over extremely long time-scales, while the Hong--Ou--Mandel and Mach--Zehnder interferometers can directly integrate the pulse-wise first-order coherence of single-photon wavepackets.

First, we will discuss the Hong--Ou--Mandel interferometer, which measures the indistinguishability of two single-photon inputs by comparing the likeness of their wavefunctions directly~\cite{Woolley2013-sz, Loudon1989-nj}. In the case of perfect identical photons at the inputs, detection of a photon by the first detector projects the second photon into the same channel as the first photon---this phenomenon is known as two-photon interference. Unfortunately, producing a single source of indistinguishable photons is often quite challenging (let alone two), but one would still like to characterize the source's instantaneous degree of first-order coherence. Or in an alternative source architecture, an individual single-photon source has recently been utilized to generate several coincident indistinguishable photons through time-multiplexing, for use in a complicated quantum-optical experiment known as Boson sampling~\cite{Loredo2016-jz, He2016-gk}. In this case, the indistinguishability of photons generated at different times, but by the same system, needs to be compared.

For these purposes, we will discuss the unbalanced and doubly-excited Mach--Zehnder interferometer. Using this interferometer, one can quantify how a source \textit{would} behave in a Hong--Ou--Mandel experiment~\cite{Muller2015-vi} by interfering pulses emitted at different times. Notably, \textit{unlike in the continuous excitation case}, the two interferometers (Hong--Ou--Mandel and Mach--Zehnder) do not necessarily measure the same interference visibilities for comparable sources, nor do they share the same threshold for nonclassicality.


\subsection{Hong--Ou--Mandel interferometer}

\begin{figure*}
\begin{indented}
\item\includegraphics{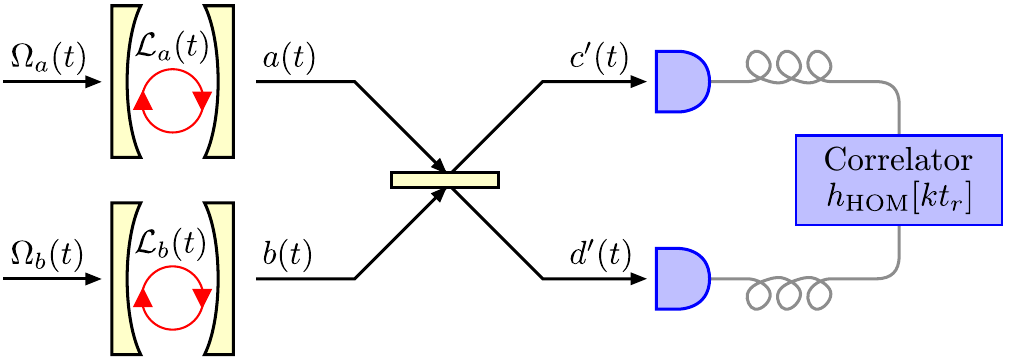}
\end{indented}
\caption{Schematic of the \textit{Hong--Ou--Mandel interferometer}: The emission from two independent sources (periodic every $t_r$) is interfered on two detectors by a beamsplitter. A digital recorder then correlates the detection times and computes $h_{\scriptsize\textrm{HOM}}[kt_r]$. Here, $\Omega_a(t)$ and $\Omega_b(t)$ indicate the coherent driving fields, $\mathcal{L}_a(t)$ and $\mathcal{L}_b(t)$ represent the dynamics of the sources, $a(t)$ and $b(t)$ indicate the continuous mode free-field operators at the output of the driven systems, and $c'(t)$ and $d'(t)$ indicate the free-field operators at the inputs to the detectors.}
\label{figure:5}
\end{figure*}

Here, we detail interference in a Hong--Ou--Mandel (HOM) interferometer. In such an interferometer, two identically independent sources are interfered on two detectors by the action of a single beamsplitter. This setup is shown schematically in figure~\ref{figure:5}: Consider two systems periodically driven by the pulses $\Omega_a(t)$ and $\Omega_b(t)$, respectively. Their outputs, the Heisenberg free-field operators $a(t)$ and $b(t)$, are subsequently fed into the beamsplitter which mixes the two according to the unitary transformation
\begin{equation}
\left[ \begin{array}{c}
c(t) \\
d(t) \\
\end{array} \right]
=
\frac{1}{\sqrt{2}}
\left[ \begin{array}{cc}
1 & -1 \\
1 & 1 \\
\end{array} \right]
\left[ \begin{array}{c}
a(t) \\
b(t) \\
\end{array} \right]\textrm{.}
\end{equation}
The detection events are then correlated electronically to arrive at a temporal coincidence histogram
\begin{equation}
h_{\scriptsize\textrm{HOM}}[kt_r]=\sum_{n=0}^{\lfloor\frac{T_{\scriptsize\textrm{int}}}{t_r}\rfloor} m_{c'}[nt_r]m_{d'}[(n+k)t_r]\textrm{.}
\end{equation}
In terms of the instantaneous correlations discussed in section~\ref{sec:inst_corr}, we wish to calculate $G^{(2)}_{c'd'}[0]$ which is $\propto\textrm{E}[h_{\scriptsize\textrm{HOM}}[0]]$. To perform this calculation, we first decomposed the underlying instantaneous correlations of $G^{(2)}_{c'd'}[0]$, i.e. $G^{(2)}_{c'd'}\left(t, t' \right)$, into more manageable components based on $a(t)$ and $b(t)$ that allowed for the averaging over all quickly-varying phase terms. These phase terms are difficult to observe experimentally since they depend on femtosecond phase locking and are, regardless, not required for the observation of two-photon interference.

It has been shown that after performing this procedure~\cite{Woolley2013-sz} one arrives at
\begin{equation}
\hspace{-30pt}
\eqalign{G^{(2)}_{c'd'}\left(t, t' \right) =
\frac{1}{4} \sum_{(e,f)\in\prod(a,b)}
\Big[ G^{(2)}_{ee} \left(t, t' \right) + G^{(1)}_{e} \left(t, t \right)G^{(1)}_{f} \left(t', t' \right) -  \\  \hspace{240pt}
\left[G^{(1)}_{e} \left(t, t' \right)\right]^* G^{(1)}_{f} \left(t, t' \right) \Big]}\textrm{,}\label{eq:42}
\end{equation}
where $G^{(1)}_e \left(t, t' \right)=\langle e^{\dagger}(t)e(t') \rangle$ represents the sources' first-order optical coherences and $\prod(a,b)\equiv\{(a,b),(b,a)\}$. Then we again remark that in its pulse-wise form
\begin{equation}
G^{(2)}_{c'd'}[0]=\int_0^T \int_0^T \mathop{\textrm{d} t} \mathop{\textrm{d} t'} \, G^{(2)}_{c'd'}\left(t, t' \right)\textrm{.}\label{eq:43}
\end{equation}
This formulation can easily represent any form of inter-source distinguishability, possibly arising due to a frequency difference, temporal or amplitude mode mismatch, multiphoton pulse, or an excitation timing jitter. However, we need a valid intensity reference to compare this correlation against in experiment due to unknown detection and collection efficiencies (see the Appendix for a detailed discussion on this normalization choice). Similar to the long-time correlation in the HBT setup, here the only true reference is in proportion to the statistically independent cross-correlation
\begin{equation}
\lim_{k\rightarrow\infty}G^{(2)}_{c'd'}[kt_r]= \frac{1}{4}
\left(\langle \hat{M}_a(T) \rangle + \langle \hat{M}_b(T) \rangle\right)^2
\textrm{.}\label{eq:44}
\end{equation}
Since we're only interested in determining the pulse-wise correlations, we consider
\begin{equation}
g^{(2)}_\textrm{\scriptsize HOM}[0] \equiv g^{(2)}_{c'd'}[0]=
\lim_{k\rightarrow\infty}\frac{G^{(2)}_{c'd'}[0]}
{G^{(2)}_{c'd'}[kt_r]}=
\frac{4G^{(2)}_{c'd'}[0]}
{\left(\langle \hat{M}_a(T) \rangle + \langle \hat{M}_b(T) \rangle\right)^2}
\label{eq:g2HOM_general}
\end{equation}
whose estimate from the HOM setup is
\begin{equation}
\hat{g}^{(2)}_\textrm{\scriptsize HOM}[0] =\lim_{k\rightarrow\infty}\frac{h_{\scriptsize\textrm{HOM}}[0]}{
h_{\scriptsize\textrm{HOM}}[kt_r]}\left(1\pm\sqrt{\frac{1}{h_{\scriptsize\textrm{HOM}}[0]}+\frac{1}{h_{\scriptsize\textrm{HOM}}[kt_r]}}\right)\textrm{.}
\end{equation}

While the analytical expression for $g^{(2)}_\textrm{\scriptsize HOM}[0]$ is somewhat complicated once expanded, one can derive insight by investigating limiting cases. For instance, consider two (potentially distinguishable) single-photon sources as the inputs to the HOM interferometer. Then, 
\begin{equation}
g^{(2)}_\textrm{\scriptsize HOM}[0] = \frac{1}{2}\left(1- \textrm{Re}
\int_0^T \int_0^T \mathop{\textrm{d} t} \mathop{\textrm{d} t'} \,
\left[G^{(1)}_{a} \left(t, t' \right)\right]^* G^{(1)}_{b} \left(t, t' \right)
\right)\textrm{,}
\end{equation}
which could represent the interference between two single-photon sources with different temporal delays, center frequencies, or pulse shapes. This expression can be understood as measuring the overlap of the sources' single-photon wavefunctions: When the sources perfectly share the aforementioned characteristics, then they have identical first-order coherences. When the first-order coherences match, so do the wavefunctions and thus $g^{(2)}_\textrm{\scriptsize HOM}[0]=0$ is measured~\cite{Loudon1989-nj}. Importantly, we now have the second requirement for a good single-photon source---\textit{a good source must be able to interfere with other sources, which requires both the wavepackets to be identical and hence in pure states}~\cite{Woolley2013-sz, Loudon1989-nj}. This is met only if both states have complete first-order coherence. Unlike for the HBT interferometer, it is also this dependence on the first-order coherence that makes Monte--Carlo simulation of the HOM histogram much more difficult.

Finally, in preparation for our comparison with the Mach--Zehnder interferometer, we look to make a few more simplifications. We note that if the sources are both identically independent and temporally matched, but we allow for non-ideal second-order coherence, then equation~(\ref{eq:g2HOM_general}) simplifies significantly to
\begin{equation}
\hspace{-20pt}
G^{(2)}_{c'd'}\left(t_1, t_2 \right) = \frac{1}{2} \left[ G^{(2)}_{aa} \left(t_1, t_2 \right) +  G^{(1)}_a \left(t_1, t_1 \right)G^{(1)}_a \left(t_2, t_2 \right) - \left| G^{(1)}_a \left(t_1, t_2 \right) \right|^2 \right]\textrm{.}
\label{eq:G2HOM_identical_source}
\end{equation}
Here, the source correlations between $c'(t)$ and $d'(t)$ have terms only coming from a single source (the indices again could trivially be switched as $a(t)$ and $b(t)$ are identically independent). Now the proportionality of the square of the experimental intensity reference reduces to
\begin{equation}
\lim_{k\rightarrow\infty}G^{(2)}_{c'd'}[kt_r]= 
\langle \hat{M}_a(T) \rangle^2
\textrm{.}
\end{equation}
Therefore, we are interested in the computation of the integrated and normalized correlation
\begin{equation}
g^{(2)}_\textrm{\scriptsize HOM}[0] = \frac{1}{2} \left[g^{(2)}_{aa} [0] + 1 - \left|g^{(1)}_a [0]\right|^2\right]\textrm{,} \label{eq:g2hom}
\end{equation}
where
\begin{equation}
|g_a^{(1)}[0]|^2\equiv\frac{|G_a^{(1)}[0]|^2}{\langle\hat{M}_a(T)\rangle^2} \quad\textrm{and} \quad |G_a^{(1)}[0]|^2\equiv\int_0^T\int_0^T\mathop{\textrm{d}t}\mathop{\textrm{d}t'}|G_a^{(1)}(t,t')|^2.
\end{equation}
One may immediately notice the nonclassical threshold of $g^{(2)}_\textrm{\scriptsize HOM}[0]<0.5$ due to the classical limit of $g^{(2)}_{aa}[0]\geq1$. Further, for a single-photon source with no error rate, $|g_a^{(1)}[0]|^2$ is precisely the trace purity of the single-photon emission. But given a finite error rate, this equivalence no longer holds. To see this, consider the resonantly driven two-level system with no dephasing: the emitted state is a pure state (with unity trace purity) even though $|g_a^{(1)}[0]|^2<1$.

Considering Eq. \ref{eq:g2hom}, the observable pulse-wise HOM interference for identically independent sources simply comprises terms involving the total first- and second-order coherences of the source. As with the continuous form of HOM interference~\cite{Woolley2013-sz}, whether this pulse-wise interference is truly just two-photon interference therefore depends on the nature of the source outputs. Additionally, we note that in many real systems the first-order coherence extracted here is wildly different than the first-order coherence extracted from a Michelson interferometer due to the long-time averaging action of the Michelson. 


\subsection{Unbalanced and doubly-excited Mach--Zehnder interferometer}

\begin{figure*}
\begin{indented}
\item\includegraphics{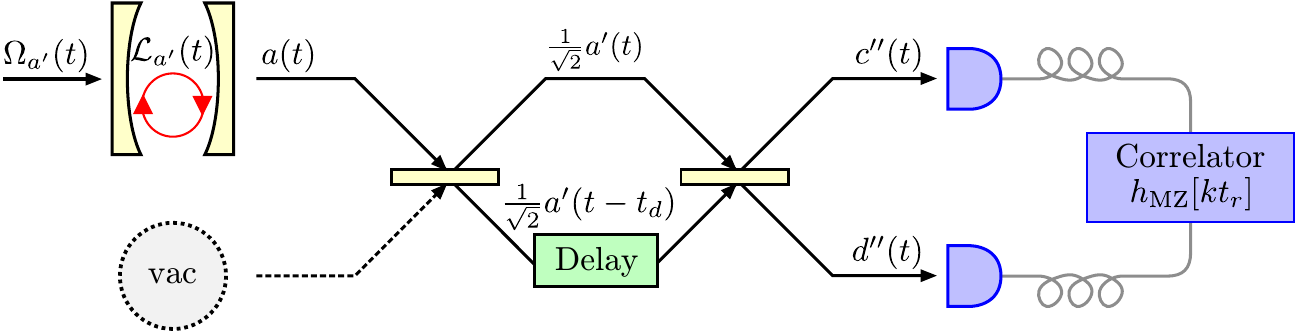}
\end{indented}
\caption{Schematic of the \textit{unbalanced Mach--Zehnder interferometer}: A single source is doubly excited at an interval of $t_d=1.9\,\textrm{ns}$ periodically (every $t_r$), to have its emission interfered with a time-delayed copy of itself. Importantly, this re-excitation only occurs after all excited population from the first pulse has decayed, so that the operator $a'(t)$ is identically independent between excitations. The vacuum mode operator at the input to the first beamsplitter has been omitted in anticipation that our detectors only measure a normally-ordered moment of their impinging fields. A digital recorder then again correlates the detection times and computes $h_{\scriptsize\textrm{MZ}}[kt_r]$. Here, $\Omega'_a(t)=\Omega_a(t) + \Omega_a(t-t_d)$ indicates the coherent driving field, $\mathcal{L}_{a'}(t)$ represents the dynamics of the source, $a'(t)=a(t)+a(t-t_d)$ indicates the continuous mode free-field operator at the output of the driven system, and $c''(t)$ and $d''(t)$ indicate the free-field operators at the inputs to the detectors.}
\label{figure:6}
\end{figure*}

While the HOM interferometer compares two single photons between a pair of different sources, the goal of the Mach--Zehnder (MZ) interferometer (shown schematically in figure~\ref{figure:6}) is to ascertain how a single source \textit{would} behave were it and its identical twin fed into an HOM interferometer. To this end, it was previously realized that some aspect of two-photon interference can be observed for a single source at the input to a MZ interferometer~\cite{Santori2002-hd}. However, we note that further formalism was required beyond previous analyses, which we will provide here.

In order to observe this type of interference, the system must be doubly excited at a time interval that matches the temporal delay between the two paths of the MZ interferometer. Additionally, this interval must be long enough that the emission resulting from the two excitations is identically independent. For perfect single-photon input the MZ output correlation is in fact two-photon interference and thus equivalent to HOM interference. However, in many cases the literature has incorrectly compared the two when the source has some finite probability of emitting non-single photons. Thus, we now outline the derivation of the correct measured correlation. Consider the schematic in figure~\ref{figure:6}: A single system is periodically driven with the coherent driving field $\Omega_a'(t)=\Omega_a(t) + \Omega_a(t-t_d)$ and has field-mode operator output $a'(t)=a(t)+a(t-t_d)$, where $t_d$ is long enough so that the state of the system is reset between excitations. The first beamsplitter simply splits $a'(t)$ in two---we have dropped the vacuum operator in anticipation that our detectors only measure a normally-ordered moment of their impinging fields. Next, one path is delayed by the excitation interval $t_d$ and the second beamsplitter then mixes $\frac{1}{\sqrt{2}}a'(t)$ with the time delayed version of itself. The time delay is absolutely critical: For time delays much less than $t_d$, $\frac{1}{\sqrt{2}}a'(t)$ can interfere with itself, and therefore reproduce a correlation similar to the HOM cross-correlation. Finally, the detection apparatus generates the histogram
\begin{equation}
h_{\scriptsize\textrm{MZ}}[kt_r]=\sum_{n=0}^{\lfloor\frac{T_{\scriptsize\textrm{int}}}{t_r}\rfloor} m_{c''}[nt_r]m_{d''}[(n+k)t_r]\textrm{.}
\end{equation}

Following the methods outlined in the previous section, we similarly calculate $G^{(2)}_{c''d''}[0]\propto\textrm{E}[h_{\scriptsize\textrm{MZ}}[0]]$ in terms of the instantaneous correlations. Although computation of $G^{(2)}_{c''d''}[0]$ in terms of source correlations certainly appears daunting due to the doubly excited source, the independence of $a'(t)$ for times of order $t_d$ or larger dramatically simplifies calculation of $G^{(2)}_{c''d''}[0]$. This expansion is further simplified by consideration of the correlations centered around time delays that are integer multiples of $t_d$. It is fairly trivial to show that the majority of terms for delays larger than $t_d$ are zero, with only a few additional phase-dependent terms arising. We note that while one might expect the correlations to be phase locked by the MZ interferometer, typical integration times quickly destroy the phase interference; 
Besides, this phase dependence is unwanted for two-photon interference~\cite{Woolley2013-sz}.

Consider the correlations about zero delay: $a'(t)$ and its time delayed version are statistically independent so equation~(\ref{eq:G2HOM_identical_source}) holds but with the source correlations of $a'(t)$ instead of $a(t)$. Because the source is doubly excited the ratio of the correlations in terms of $a(t)$ is altered, which can easily be seen by applying the above rules. Now,
\begin{equation}
G^{(2)}_{c''d''}[0] = 2G^{(2)}_{aa} [0] + \left[\langle \hat{M}_a(t) \rangle^2 - \left| G^{(1)}_a [0] \right|^2 \right]\textrm{.}
\end{equation}
Again, we need a long-time reference for the square of the experimental intensity, which is proportional to
\begin{equation}
\lim_{k\rightarrow\infty}G^{(2)}_{c''d''}[kt_r]= 3\langle \hat{M}_a(t) \rangle^2
\end{equation}
and so we consider the normalized pulse-wise correlation
\begin{equation}
g^{(2)}_\textrm{\scriptsize MZ}[0]\equiv g^{(2)}_{c''d''}[0]=
\lim_{k\rightarrow\infty}\frac{G^{(2)}_{c''d''}[0]
}
{G^{(2)}_{c''d''}[kt_r]}\textrm{.}
\end{equation}
Thus, we arrive at
\begin{equation}
g^{(2)}_\textrm{\scriptsize MZ}[0] = \frac{2}{3}g^{(2)}_{aa} [0] + \frac{1}{3} \left[ 1 - \left| g^{(1)}_a [0] \right|^2 \right]\textrm{,}
\end{equation}
whose estimate from the MZ setup is given by
\begin{equation}
\hat{g}^{(2)}_\textrm{\scriptsize MZ}[0] =\lim_{k\rightarrow\infty}\frac{h_{\scriptsize\textrm{MZ}}[0]}{
h_{\scriptsize\textrm{MZ}}[kt_r]}\left(1\pm\sqrt{\frac{1}{h_{\scriptsize\textrm{{MZ}}}[0]}+\frac{1}{h_{\scriptsize\textrm{{MZ}}}[kt_r]}}\right)\textrm{.}
\end{equation}
Importantly, we note that this different interference visibility as compared with equation~(\ref{eq:g2hom}) for the HOM setup holds only for the pulsed case~\cite{Muller2015-om}. Additionally, the threshold for a nonclassical correlation has now changed to $g^{(2)}_\textrm{\scriptsize MZ}[0]<\frac{2}{3}$. On the other hand, if the MZ interferometer is used for two-photon interference under continuous excitation, its interference visibility and nonclassical threshold actually match that of the HOM interferometer.

\begin{figure}
\begin{indented}
\item\includegraphics{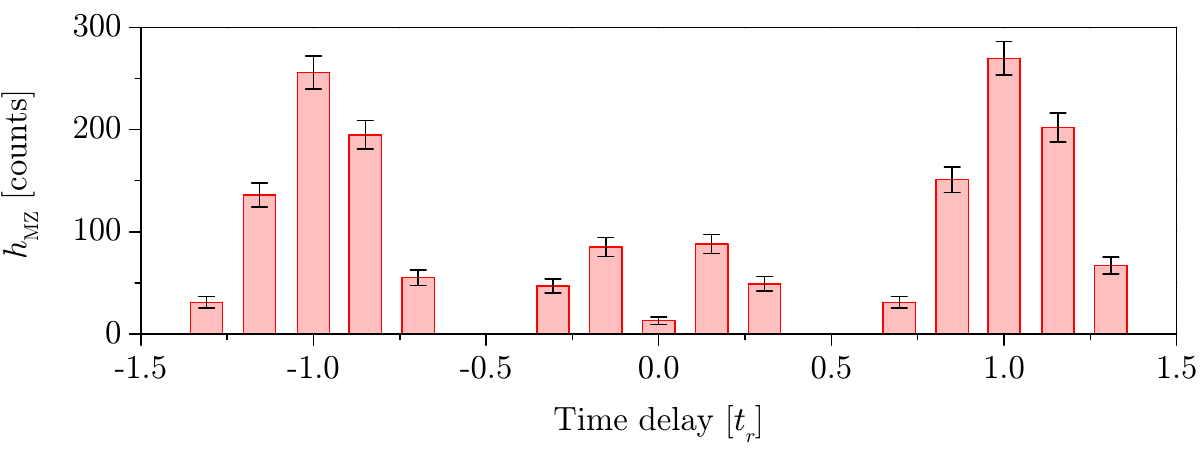}
\end{indented}
\caption{Example unbalanced and doubly-excited Mach--Zehnder histogram generated on pulsed single-photon emission from a cavity quantum electrodynamical system. An estimate for the Hong--Ou--Mandel interference that would occur, should the source's identical copy and itself be input to an Hong--Ou--Mandel interferometer, can be obtained by fitting the characteristic five-peaked distribution. This distribution contains enough information to accurately determine both the degrees of first- and second-order optical coherence.}
\label{figure:7}
\end{figure}

Finally as an experimental example, we discuss the observed two-photon interference in a MZ setup when interfering single-photon generation from a cavity-quantum electrodynamical source (figure~\ref{figure:7}); The data is reproduced from reference~\cite{Muller2015-om}. The MZ histogram shows groups of five peaks with non-trivial correlations. By applying the above analysis to each delay, i.e. $\tau\in{\{-2t_d,-t_d,0,t_d,2t_d\}}$, we can compute $g^{(2)}_\textrm{\scriptsize MZ}[kt_d]$ in terms of source correlations. However, the experimental beamsplitters often have a transmittivity ($T$) to reflectivity ($R$) ratio deviating from $50:50$. Therefore, to fit the observed five-peak pattern in correlation measurements, we need to include the individual transmittivities and reflectivities of the first beamsplitter ($T_1, R_1$) and second beamsplitter ($T_2, R_2$). Recently, a very intuitive set of pictorial rules of this process were given in the supplementary information of reference~\cite{Loredo2016-zs}. Combining these rules with our formalism, the amplitudes of the five peaks in correlation measurements are given by:

\begin{equation}
\hspace{-70pt}
\label{cases}
g^{(2)}_\textrm{\scriptsize MZ}[nt_d] = \cases{
\frac{8}{3}R_1 T_1 R_2^2&if $n=-2$\\
\frac{8}{3}R_2T_2\left(R_1^2+T_1^2\right) + \frac{16}{3} g^{(2)}_a[0] R_1T_1R_2^2&if $n=-1$\\
\frac{16}{3} g^{(2)}_a[0] R_2T_2\left(R_1^2+T_1^2\right) + \frac{8}{3}R_1T_1 \left(R_2^2+T_2^2 - 2|g^{(1)}_a[0]|^2R_2T_2\right)&if $n=0$\\
\frac{8}{3}R_2T_2\left(R_1^2+T_2^2\right) + \frac{16}{3} g^{(2)}_a[0] R_1T_1T_2^2&if $n=1$\\
\frac{8}{3}R_1 T_1 T_2^2&if $n=2$\\}\textrm{.}
\end{equation}
%


\subsection{Simulated first-order optical coherences of single-photon sources and hence two-photon interference\label{sec:sim2}}

In the previous simulation section~\ref{sec:sim1}, we examined the second-order optical coherences of our three prototypical systems. We saw that under the right conditions, each system could act as a single-photon source that emitted precisely only one photon per pulse. However in this simulation section, we will discuss the other requirement of an ideal single-photon source, that it possess complete first-order optical coherence. We will calculate the expected first-order coherence for each of our model systems, and will directly show the expected two-photon interference between photons of the prototypical systems.

\subsubsection{Identical inputs}
\begin{figure}
\begin{indented}
\item\includegraphics{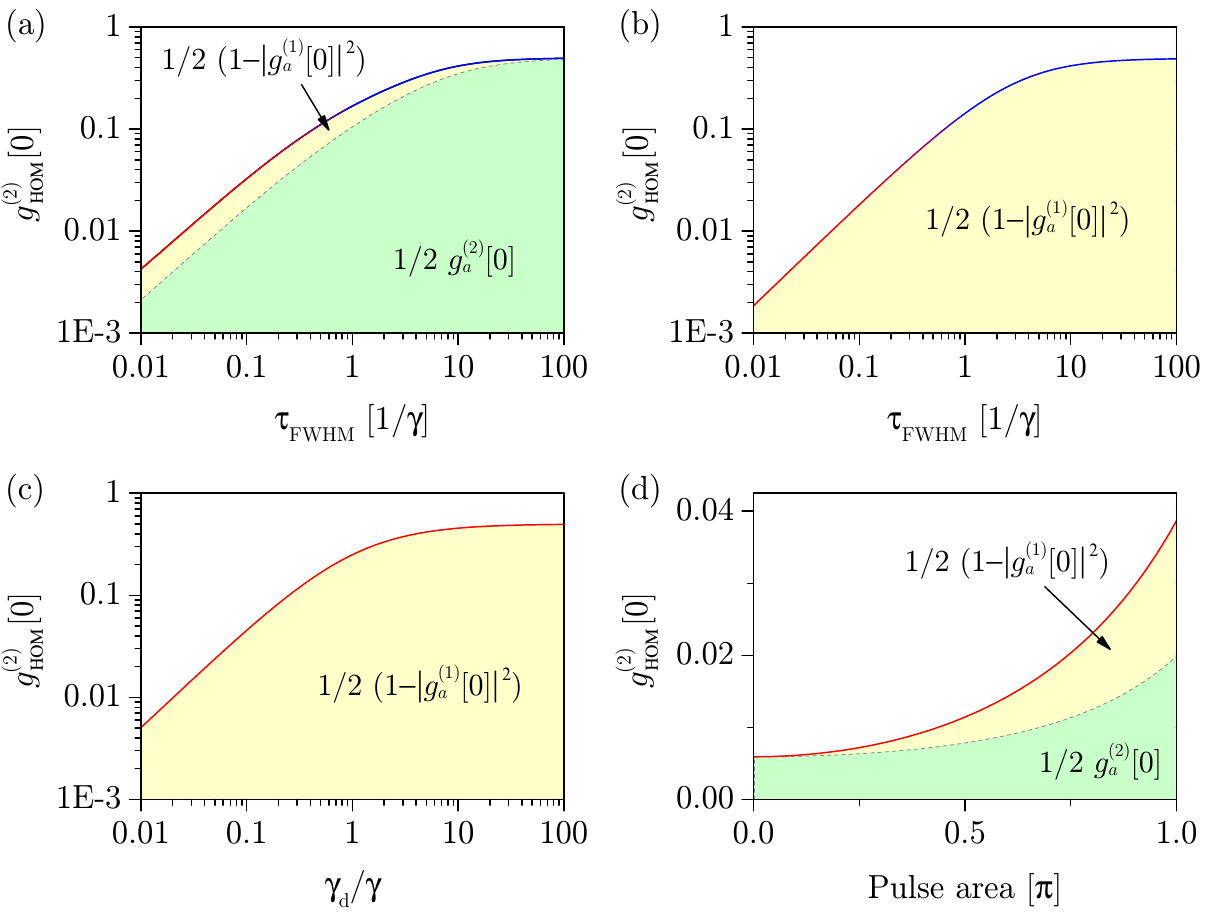}
\end{indented}
\caption{Simulations of the measured-degrees of second-order Hong--Ou--Mandel coherence $g^{(2)}_{\scriptsize\textrm{HOM}}[0]$, for the three representative systems highlighted in section~\ref{systems}, as if the emission from two identical systems impinged on the inputs. $g^{(2)}_{\scriptsize\textrm{HOM}}[0]$ is graphically broken into two colors to depict portions dependent on the source's first-order (yellow shaded) or second-order (green shaded) coherences. (a) $g^{(2)}_{\scriptsize\textrm{HOM}}[0]$ as a function of excitation pulse length for system (i), showing degradation primarily in source $g^{(2)}_{aa}[0]$. (b) $g^{(2)}_{\scriptsize\textrm{HOM}}[0]$ as a function of excitation pulse length for system (ii), showing degradation exclusively in source $g^{(1)}_a[0]$. (c) $g^{(2)}_{\scriptsize\textrm{HOM}}[0]$ as a function of an additional dephasing rate for system (iii), showing degradation exclusively in source $g^{(1)}_a[0]$. Again, for the simulations in (a-c), all pulse lengths were chosen such that the average number of photons emitted is one. For (c) and (d), exponential wavepackets are generated with an excitation of pulse length $\tau_{\scriptsize\textrm{FWHM}}=0.1/\gamma$ in energy. (d) $g^{(2)}_{\scriptsize\textrm{HOM}}[0]$ as a function of excitation pulse \textit{area} for system (i), showing degradation in both source $g^{(1)}_a[0]$ and $g^{(2)}_{aa}[0]$.}
\label{figure:8}
\end{figure}
--- We discuss simulations of HOM interference for identical inputs, since the MZ interferometer configuration is very popular in characterizing single-photon sources and is closely related to a HOM experiment with identical inputs. As previously, our systems have been excited by the same Gaussian pulses and with the red (blue) lines corresponding to exponential (Gaussian) wavepacket shapes. Similarly, unless otherwise noted the pulse lengths were chosen to ensure an average of one photon per pulse. Now, we will examine the measured degrees of HOM coherence $g^{(2)}_{\scriptsize\textrm{HOM}}[0]$ (colored lines in figure~\ref{figure:8}) and their decompositions into components that depend on their sources' first-order (yellow shaded) or second-order (green shaded) coherences. 

First, again consider the pulsed two-level source [system (i)] but now with the emission from two identical systems as inputs to the interferometer. Although the emission from the two-level systems is almost exclusively incoherently scattered, it still possesses relatively good first-order coherence for all excitation pulse lengths. The first-order coherence is greatest for long pulses (at low instantaneous power) or short pulses (where no dephasing can occur): This can be seen nicely in the minimal yellow shading of figure~\ref{figure:8}a. Therefore, the $g^{(2)}_{\scriptsize\textrm{HOM}}[0]$ almost perfectly tracks the source's $\frac{1}{2}g^{(2)}_{a}[0]$ (green shaded). As with the second-order coherence, we see that emission from ideal two-level systems can only generate perfect two-photon interference within some allowable tolerance for errors. Notably, the terms involving the second- and first-order coherences degrade at identical rates for low powers. Therefore, the double-click probability has an identical lower bound as for the HBT and HOM experiments with this source, i.e.
\begin{equation}
\hat{g}^{(2)}_{\scriptsize\textrm{HOM}}[0]=0.4\gamma\,\tau_{\scriptsize\textrm{FWHM}}\pm0.003\textrm{.}
\end{equation}
For shorter excitation pulse lengths, the emission is perfectly coherent and the HOM error probability is also dominated by the second-order coherence (figure~\ref{figure:8}d).

On the other hand, interference between a pair of three-level ladder systems [systems~(ii)] perfectly tracks the degradation of first-order coherence with increasing pulse length (figure~\ref{figure:8}b). As the incoherent excitation pulse becomes longer relative to the radiative decay time, the coherence of the state coupled to the emission channel is destroyed. From a Monte--Carlo perspective, this amounts to a more randomized time of excitation of the intermediate level $|2\rangle$ and hence randomized phase of the first-order coherence. This ideal system therefore has a source error probability whose lower bound is determined by the speed with which the intermediate level can be populated.

Unlike these other systems, the ideal three-level lambda source [system (iii)] possesses complete first-order coherence for all possible excitation pulses, in addition to possessing the correct photocount distribution. However, we can manually add a non-ideality, such as a pure dephasing rate (with collapse operator $C=\sqrt{\gamma_d}\sigma_{22}$), that can destroy the HOM interference between two identical ladder systems. The HOM interference between such systems, as the dephasing rate $\gamma_d$ is increased, is shown in figure~\ref{figure:8}c. As is the case for all three scenarios in figures~\ref{figure:8}a-c, the nonclassicality of the HOM interference is completely destroyed by either a large pulse length or an explicit dephasing rate. Note: often the dephasing rates could be power dependent, such as those arising from phonon-induced dephasing~\cite{Krummheuer2002-ct}---this situation is also easily calculable from our formalism and with QuTiP. The physical effect of including a power-dependent dephasing rate is comparable to the degradation of first-order coherence seen in figures~\ref{figure:8}b~and~\ref{figure:8}c (but with increasing power), and therefore we did not include this redundant demonstration.

\subsubsection{Non-identical inputs}--- Finally, we discuss simulations of HOM interference for non-identical inputs, since sources in realistic quantum networks will most likely never be completely identical. Therefore, we believe it useful to know how to calculate interference between disparate sources. This difference may arise due to different temporal delays, center frequencies, or pulse shapes between the interfering wavepackets~\cite{Woolley2013-sz}. Because the calculational tool for achieving any of these three effects is identical, we have chosen only to discuss temporal delay and pulse shape as ways of introducing source distinguishability. These methods will be used to discuss two representative ways of destroying HOM interference.

\begin{figure}
\begin{indented}
\item\includegraphics{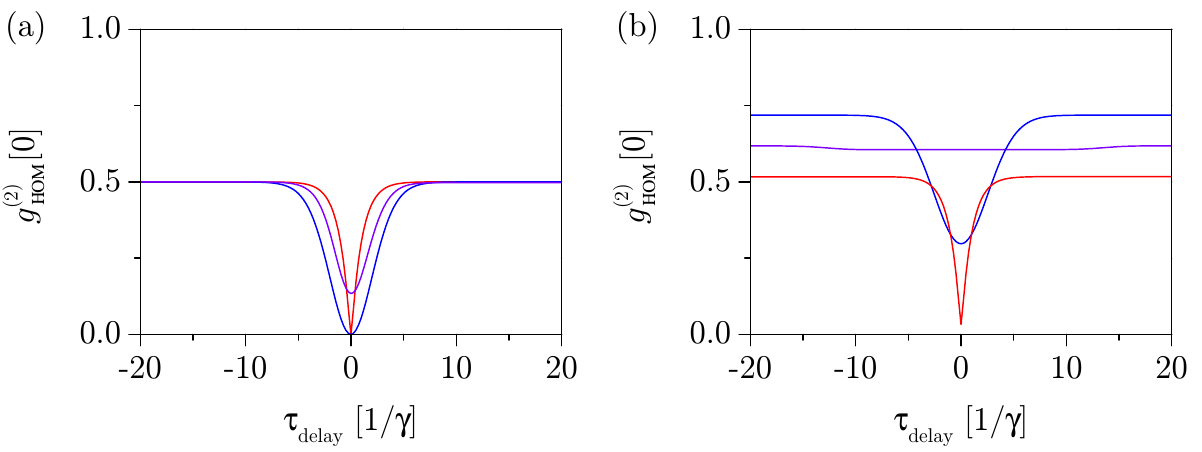}
\end{indented}
\caption{Simulations of the measured-degree of second-order Hong--Ou--Mandel coherence $g^{(2)}_{\scriptsize\textrm{HOM}}[0]$, for interference between delayed photons at the two inputs. Interference as a function of temporal delay between the emitted wavepackets is shown for photons generated by two systems of type (iii) in (a) and two systems of type (i) in~(b). Colors indicate the following types of wavepacket envelopes---Red: two exponential envelopes with pulse lengths of $\tau_{\scriptsize\textrm{FWHM}}=0.1/\gamma$ in energy, Blue: two Gaussian envelopes with pulse lengths of $\tau_{\scriptsize\textrm{FWHM}}=3.3/\gamma$ in energy, and Violet: one exponential and one Gaussian envelope with comparable pulse lengths as before. Note: the excitation pulse lengths were again chosen such that each system emitted, on average, one photon per pulse.}
\label{figure:9}
\end{figure}

For simplicity, our systems have again been excited by the same Gaussian pulses and as before, with interference between two identical exponential (Gaussian) packets denoted with the red (blue) lines. Similarly, the pulse areas were chosen to ensure an average of one photon per pulse. Now, however, we have additional purple lines that represent interference between a Gaussian wavepacket and an exponential wavepacket.

Consider HOM interference between a pair of three-level lambda systems [systems of type (iii)] centered at the same frequency but potentially excited with different pulses and temporal offsets (figure~\ref{figure:9}a). Notice that the HOM interference only occurs when the wavepackets overlap (around zero delay), while for large delays the HOM cross-correlation reaches the limit of nonclassicality ($g^{(2)}_{\scriptsize\textrm{HOM}}[0]<0.5$). Because all three-level lambda systems are perfect single-photon sources, both the red and blue curves that correspond to identical exponential or Gaussian photon inputs, respectively, exhibit perfect two-photon interference at zero delay. On the other hand, when the two wavepackets do not perfectly overlap due to their differing shapes (purple), then the interference is imperfect for any delay.

This situation is made even more non-ideal if one considers interference between wavepackets emitted by a pair of two-level systems [systems of type (i)], as shown in figure~\ref{figure:9}b. In this situation, the exponential wavepackets show good HOM interference visibility at zero delay (red). Meanwhile, in the case of the Gaussian wavepackets, the significant probability of re-excitation (due to $g^{(2)}_{a}[0]=0.44$) partially destroys the visibility of the HOM interference (blue); The HOM dip at zero delay still is below the nonclassical threshold. However, if Gaussian and exponential wavepackets, each generated by two-level systems, are interfered with one-another then the HOM interference is not below the nonclassical threshold (purple). Interestingly, the destruction of the HOM visibility is much worse for the wavepackets generated by the two-level systems as compared to that generated by the three-level lambda systems. This is a result of the re-excitation action that scrambles the phase of the first-order coherence. Therefore, only a minimal first-order contribution to the HOM interference occurs over the entire range of roughly $\tau_{\scriptsize\textrm{delay}}\pm10/\gamma$.


\section{Conclusions}

Herein, we have thoroughly discussed the properties of on-demand pulsed single-photon sources. In particular, we provided a general recipe for the complete numerical characterization of the behavior of pulsed single-photon sources. Using our technique, we looked beyond previous studies in order to completely describe the non-idealities of single-photon sources. By considering the character of single-photon emission from three prototypical systems, we believe we have provided a set of source behaviors that most other pulsed single-photon sources can be mapped onto. In this way, we hope that our work will be used to understand more complex single-photon sources and establish lower-bounds for potential source non-idealities. Considering its applicability to more complicated systems, we have already used these techniques to characterize a cavity quantum electrodynamical source~\cite{Muller2015-vi, Muller2015-il, Muller2015-om}.

Through detailed simulations that utilized an extended form of the quantum regression theorem, we showed how such sources are characterized by the following attributes:

\begin{itemize}
\item \textit{Photocount distribution}, yielding at most one photon per pulse and hence zero \textit{second-order optical coherence},
\item \textit{First-order optical coherence}, such that the output pulse is a pure state of the free radiation field rather than an incoherent mixture of single-photon pulses,
\item \textit{Spatio-temporal profile}, whose precise shape is tailored by the generating system and usefulness determined by its application.
\end{itemize}

The photocount distribution and second-order coherence can be observed in the Hanbury--Brown and Twiss setup, while the first-order optical coherence manifests itself experimentally in Hong--Ou--Mandel and Mach--Zehnder interferometers. Finally, we have worked to make these characterizations easily accessible to experimentalists wishing to model their on-demand sources by contributing to the underlying code of the open-source package Quantum Toolbox in Python (QuTiP). In addition, we have provided detailed examples to the QuTiP repository demonstrating many of the simulations discussed in this paper.


\section{Acknowledgements}

The authors thank Yousif Kelaita and Vinay Ramasesh for productive and helpful discussions in framing the context of this work.

We gratefully acknowledge financial support from the National Science Foundation (Division of Materials Research - Grant. No. 1503759). KAF acknowledges support from the Lu Stanford Graduate Fellowship and the National Defense Science and Engineering Graduate Fellowship. KM acknowledges support from the Alexander von Humboldt Foundation.



\addcontentsline{toc}{section}{Appendix: Normalization of the Hong--Ou--Mandel and Mach--Zehnder correlations}
\section*{Appendix: Normalization of the Hong--Ou--Mandel and Mach--Zehnder correlations}

First, we briefly describe how to arrive from Eq. \ref{eq:42} to \ref{eq:44} in the main text. Consider Eq. \ref{eq:42} in the limit
\begin{equation}
\nonumber\eqalign{\lim_{k\rightarrow\infty}G^{(2)}_{c'd'}\left(t, t'+kt_r \right) =\\\quad\nonumber
\frac{1}{4} \sum_{(e,f)\in\prod(a,b)}
\Big[ G^{(2)}_{ee} \left(t, t' +kt_r\right) + G^{(1)}_{e} \left(t, t \right)G^{(1)}_{f} \left(t'+kt_r, t'+kt_r \right) -  \nonumber\\  \hspace{140pt}
\left[G^{(1)}_{e} \left(t, t' +kt_r\right)\right]^* G^{(1)}_{f} \left(t, t'+kt_r \right) \Big]\nonumber.}\nonumber\label{eq:60}
\end{equation}
Note that the first-order coherence inherits the envelope of coherence decay from the excitation laser for long times, and hence always vanishes in the long time limit
\begin{eqnarray}
\lim_{k\rightarrow\infty}G^{(1)}_{e} \left(t, t'+kt_r \right)=0.
\end{eqnarray}
This is an experimental consideration that may be difficult in observing due to various long-time effects such as blinking, potentially limited laser coherence, or operating the correlator in a start-stop configuration (see the main text). In the long time limit, the second-order coherence terms in Eq.~\ref{eq:60} become uncorrelated
\begin{eqnarray}
\lim_{k\rightarrow\infty}G^{(2)}_{ee} \left(t, t'+kt_r \right)&=& \langle: e^\dagger(t)e(t)e^\dagger (t'+kt_r)e(t'+kt_r):\rangle\\
&=&\langle e^\dagger (t)e(t)\rangle\langle e^\dagger (t'+kt_r)e(t'+kt_r)\rangle.
\end{eqnarray}
 Then, using the fact that every pulse period is identical
\begin{eqnarray}
\lim_{k\rightarrow\infty}G^{(2)}_{ee} \left(t,t'+ kt_r \right)&=&  G^{(1)}_{e} \left(t, t \right) G^{(1)}_{e} \left(t', t' \right),
\end{eqnarray}
and also the photon flux terms in Eq. \ref{eq:60} have a similar point
\begin{eqnarray}
\lim_{k\rightarrow\infty}G^{(1)}_{f} \left(t'+kt_r, t' +kt_r\right)=G^{(1)}_{f} \left(t', t' \right).
\end{eqnarray}
As a result, we can write
\begin{eqnarray}
\lim_{k\rightarrow\infty}G^{(2)}_{c'd'}\left(t, t'+kt_r \right) =\\\nonumber
\qquad\frac{1}{4} \sum_{(e,f)\in\prod(a,b)}
 G^{(1)}_{e} \left(t, t \right) G^{(1)}_{e} \left(t', t' \right) + G^{(1)}_{e} \left(t, t \right)G^{(1)}_{f} \left(t', t' \right).
\end{eqnarray}
Integrating to pulse-wise form with Eq. \ref{eq:43} of the paper
\begin{equation}
G^{(2)}_{c'd'}[0]=\int_0^T \int_0^T \mathop{\textrm{d} t} \mathop{\textrm{d} t'} \, G^{(2)}_{c'd'}\left(t, t' \right)\textrm{,}
\end{equation}
and using the definition $\langle \hat{M}_e(T) \rangle=\int_0^T\mathop{\textrm{d} t} \, G^{(1)}_{e}\left(t, t \right)$, we have the result of Eq. \ref{eq:44} in the paper
\begin{equation}
\lim_{k\rightarrow\infty}G^{(2)}_{c'd'}[kt_r]= \frac{1}{4}
\left(\langle \hat{M}_a(T) \rangle + \langle \hat{M}_b(T) \rangle\right)^2
\textrm{.}
\end{equation}

The potentially confusing point of this definition is that for a good single-photon source
\begin{equation}
\lim_{k\rightarrow\infty}G^{(2)}_{c'd'}[kt_r]\neq G^{(2)}_{c'd'}[t_r]
\end{equation}
because the first-order coherence terms in Eq. \ref{eq:60} interfere with the intensity for short times resulting in
\begin{equation}
G^{(2)}_{c'd'}[t_r]\approx \frac{1}{4}
\left(\langle \hat{M}_a(T) \rangle^2 + \langle \hat{M}_b(T) \rangle^2\right).\label{eq:70}
\end{equation}
Hence, using the normalization $G^{(2)}_{c'd'}[0]/G^{(2)}_{c'd'}[t_r]$ results in a denominator that depends on the first-order coherence, which we believe is not the most ideal definition. An alternative way of achieving our preferred normalization is to use the average of the cross- and auto-correlations
\begin{equation}
g^{(2)}_{c'd'}[0]=\lim_{k\rightarrow\infty}\frac{G^{(2)}_{c'd'}[0]}{G^{(2)}_{c'd'}[kt_r]}=\frac{4G^{(2)}_{c'd'}[0]}{2G^{(2)}_{c'd'}[t_r]+G^{(2)}_{c'c'}[t_r]+G^{(2)}_{d'd'}[t_r]},
\end{equation}
where 
\begin{equation}
2G^{(2)}_{c'd'}[t_r]+G^{(2)}_{c'c'}[t_r]+G^{(2)}_{d'd'}[t_r]= \frac{1}{4}
\left(\langle \hat{M}_a(T) \rangle + \langle \hat{M}_b(T) \rangle\right)^2.
\end{equation}
Here, we used the fact that $g^{(2)}_{c'd'}[kt_r]=g^{(2)}_{d'c'}[kt_r]$, and assumed that there are no blinking effects. Another common experimental trick to get a normalization by the intensity in Eq. \ref{eq:70} is to introduce distinguishability (e.g. via polarization rotation) between the two sources so that the fields cannot interfere at the detectors.


\addcontentsline{toc}{section}{References}
\section*{References}

\bibliographystyle{iopart-num}
\bibliography{Dynamical_two-photon_interference_v9}

\end{document}